# Atomically-thin quantum dots integrated with lithium niobate photonic chips

Daniel White[1], Artur Branny[1], Robert J. Chapman[2], Raphaël Picard[1], Mauro Brotons-Gisbert[1], Andreas Boes[2], Alberto Peruzzo[2], Cristian Bonato[1], Brian D. Gerardot[1,*]

[1]Institute of Photonics and Quantum Sciences, SUPA, Heriot-Watt University, Edinburgh EH14 4AS, UK
[2] Quantum Photonics Laboratory and Centre for Quantum Computation and Communication Technology, School of Engineering, RMIT University, Melbourne, Victoria 3000, Australia
*b.d.geradot@hw.ac.uk

The electro-optic, acousto-optic and nonlinear properties of lithium niobate make it a highly versatile material platform for integrated quantum photonic circuits. A prerequisite for quantum technology applications is the ability to efficiently integrate single photon sources, and to guide the generated photons through ad-hoc circuits. Here we report the integration of quantum dots in monolayer WSe$_2$ into a Ti in-diffused lithium niobate directional coupler. We investigate the coupling of individual quantum dots to the waveguide mode, their spatial overlap, and the overall efficiency of the hybrid-integrated photonic circuit.

## INTRODUCTION

Solid-state quantum emitters have emerged in the past few years as prominent systems for quantum technology. Single-photon sources based on semiconductor quantum dots [1], defects [2] and impurities [3] have reached extremely high purity [4] and efficiency [1], making them competitive for future quantum applications such as quantum communication [5] or linear optical quantum computing [6]. Additionally, spin-active quantum emitters provide an excellent platform for quantum networking, enabling efficient interfacing between photons to share quantum states over long distances, and spins, which can store and process quantum information locally [7], [8].

When implementing either purely optical or hybrid spin-photon schemes, a large number of optical components are required to construct the required photonic quantum gates. For increasing number of qubits, approaches based on bulk optics become quickly impractical, due to the size and the inherent instability of these individual components. Only photonic integration can provide the stability, repeatability and robustness required for high-visibility quantum interference and scalability [9]. For these reasons a large research effort in the past decade has been dedicated to the development of integrated quantum photonics circuits with embedded quantum emitters [10]–[13]. However, several challenges are still outstanding, such as total internal reflection in bulk high-index materials that severely limits optical collection efficiency and the in-situ manipulation of the phase in interferometer arms. Furthermore, materials such as diamond, while featuring excellent spin properties, are notoriously hard to grow and fabricate, posing a serious challenge to the development of a large-scale integrated platform.

In this respect, quantum dots in two-dimensional layered materials such as transition metal dichalcogenide (TMDs) exhibit promising properties. Bulk TMDs form layered structures with weak interlayer van-der-Waals interactions which can be micro-mechanically exfoliated to obtain monolayer crystals [14]. One of the most prominent examples is WSe$_2$ which, at the monolayer level, is a direct band-gap semiconductor featuring bright localized excitons [15] or bi-excitons [16] at cryogenic temperatures. Once exfoliated, the TMD material can be stacked to create layer-by-layer van der Waals heterostructures with designer functionality [17], [18] or placed onto arbitrary substrates [19], including complex photonic circuits, using a variety of transfer methods [20]. As quantum emission is embedded in the single atomic layer, optical extraction is not limited by total internal reflection, an enormous challenge for bulk materials such as GaAs, diamond or SiC. In addition, the monolayer can be transferred to any photonic structure fabricated in any material platform, removing any need for lattice matching or complex bonding procedures. Finally, the quantum emitter can be deterministically created at a pre-determined position by strain engineering [21]–[23].

An attractive photonic platform to integrate quantum emitters is lithium niobate. Lithium niobate is the industry standard for the photonic telecommunication industry, possessing the ideal properties for integrated photonics, including low-loss transmission in a broad wavelength range, from UV to mid-IR, and ultrafast switching capability. Crystalline lithium niobate uniquely combines nonlinear, piezoelectric, photorefractive and electro-optic effects in the same platform [24], making it an attractive system for quantum technology. Low-loss waveguides and other passive components, such as beam-splitters and ring resonators [25], have been demonstrated, both by Ti in-diffusion or ridge structures [26]. Its electro-optic properties have been exploited to build fast electrically-controlled phase modulators at gigahertz operating speeds [27], [28], with recent work reporting 210 Gbit s$^{-1}$ devices [29]. Integration with superconducting nanowire detectors, which offer single photon detection with the performance required for quantum photonics application, has already been demonstrated [30].

Here we demonstrate the coupling of quantum dots in monolayer WSe$_2$ to an integrated lithium niobate directional coupler, a fundamental building block for quantum interference experiments. The integration of 2D materials hosting quantum emitters in this mature photonic platform holds the promise to realise fully-integrated quantum circuits for future quantum technologies.

## DEVICE DESIGN, FABRICATION, AND EXPERIMENTAL SET-UP

The directional coupler was fabricated by photolithographically patterning the waveguide structure (4 µm width) into a Ti thin film (70 nm thickness) on the surface of an X-cut LiNbO$_3$ wafer. The titanium was diffused into LiNbO$_3$ in a wet oxygen atmosphere at 1010°C for several hours. After the diffusion, the wafer was diced into a 2.5 cm long chip and the chip end faces were polished to optical grade. To enable stable fiber coupling inside the cryostat, a fiber array was aligned to the waveguides and glued to one facet of the chip using cryogenic compatible epoxy glue. The coupling was found to remain constant after several cryogenic cooling cycles. The waveguide propagation loss is estimated to be 0.5 dB/cm, while the coupling efficiency from fiber to waveguide is estimated to be 3.5 dB. This is confirmed by a total loss from output fiber to input facet of 4 dB, with the reflectivity of the directional coupler measured to be 70%. The facet was confocally imaged with a 760 nm laser placed into one of the fiber output ports to illuminate the input mode. The image (see Fig. 1(a)) shows an elliptical mode with major and minor axes of 10 µm and 6 µm, respectively ($1/e^2$). The numerical aperture (NA) of the waveguide is estimated to be 0.07.

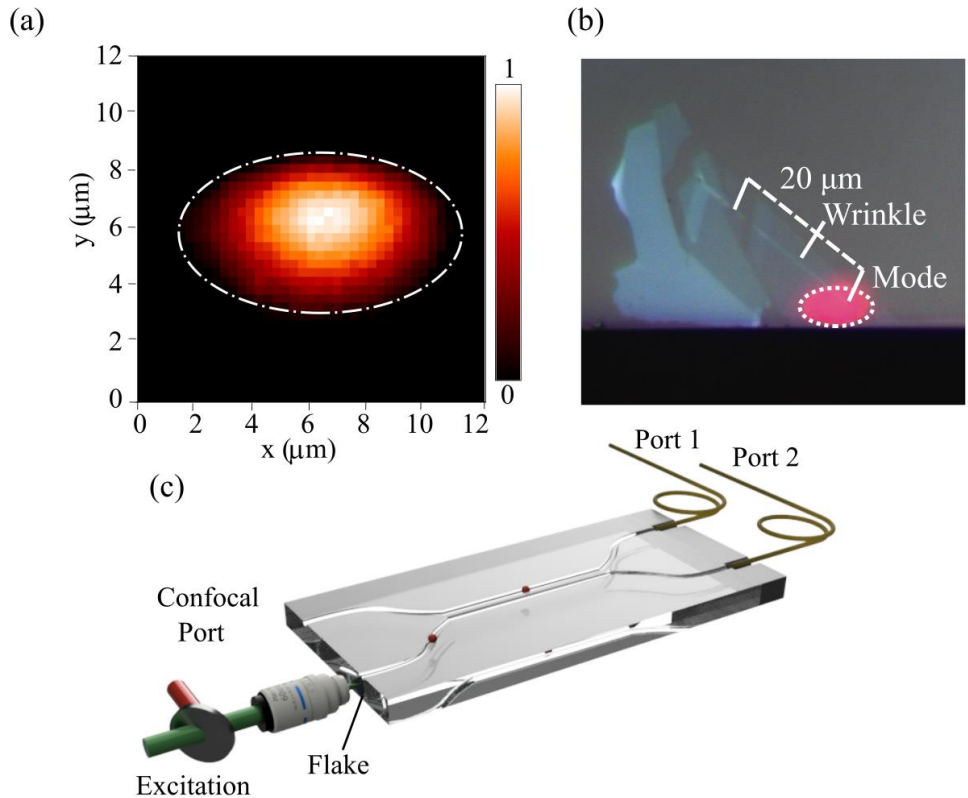

Fig. 1. (a) Spatial profile of the waveguide mode at the facet of the WSe$_2$ flake, measured by illuminating a fiber-coupled port with a 750 nm laser and detecting light through the Confocal Port. The white dash-dot line indicates the 1/e$^2$ mode profile, the color scale shows the normalized intensity. (b) Optical microscope image of WSe$_2$ transferred onto the facet of the waveguide. The 20 µm monolayer region possesses a wrinkle with localized strain. The illuminated waveguide mode (red spot) is used to align the monolayer area of the flake with respect to the optical mode. (c) Schematic of Ti in-diffused lithium niobate directional coupler with a WSe$_2$ flake at the input facet. PL was measured by exciting the emitters in a confocal microscope, while emission was detected in confocal geometry (Confocal Port) and through the two fiber output ports (Port 1 and Port 2).

Micromechanical exfoliation was used to obtain monolayer $WSe_2$ flakes, identified by optical microscopy. We select a monolayer featuring a wrinkle as it has been shown that the associated localized strain induces single photon emitters [21]. The flake was transferred onto the waveguide using the all-dry viscoelastic stamping procedure [31]. To illuminate the waveguide optical mode position, a HeNe (633nm) laser was sent from the opposite end of the directional coupler. This allowed deterministic transfer of flakes with wrinkles directly onto the optical mode. One example is shown in Fig. 1(b).

Photoluminescence (PL) measurements were performed in a confocal microscope setup with the sample mounted onto a 3-axis nanopositioner and cooled to 4K. The sample was excited off-resonantly with 1 µW of continuous wave 532 nm laser, focused to a 1µm spot through an objective of numerical aperture 0.82. The experimental setup, shown in Fig. 1(c), enables PL to be collected confocally from the $WSe_2$ flake or from the two waveguide collection ports.

## RESULTS AND DISCUSSION

The confocal image produced in Fig. 2(a) shows PL in the spectral range 720 – 735 nm from a flake transferred onto the waveguide facet. The marked areas reveal a monolayer $WSe_2$ region (solid white line) closest to the mode center and a bilayer region (dotted white line). In the bilayer region we find the characteristic indirect optical transition at around 800 nm (not shown), while the monolayer has local regions of bright emission, as shown in Fig. 2(a). In the areas of strongly localized emission, sharp spectral lines are observed over a distribution of emission wavelengths. These regions feature quantum dots and are attributed to the local strain induced by the wrinkle and the edges of the flake. To demonstrate the coupling of the quantum dot emission to the directional coupler, we spatially excite at the localized bright regions (marked by Roman numerals in Fig. 3(a, b)) and collect PL from the two fiber output ports. We find identical spectra at each output port which match the emission measured confocally, Fig. 3 (c, d). This confirms the directional coupler acts as an on-chip beam-splitter. The device operates as a 70:30 beam-splitter around the peak of the PL ensemble (720 to 760 nm). Whilst no single photon emitters were located directly at the centre of the mode (which would maximize the coupling efficiency) for this particular flake, numerous isolated emitters were still coupled to the waveguide with PL from centres visible at a displacement of up to 3.5µm.

An estimate of the coupling efficiency can be obtained by calculating the dipole emission collected by a waveguide

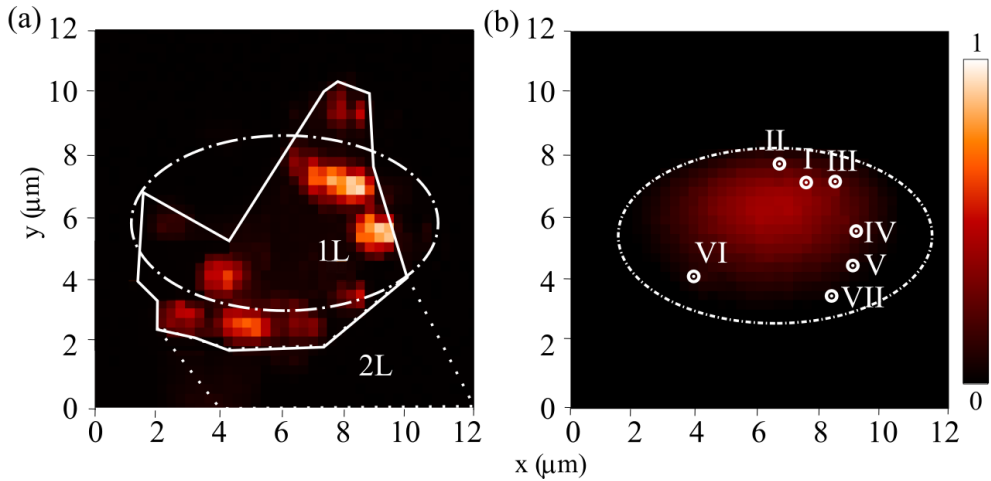

Fig. 2. (a) Confocal PL image of the $WSe_2$ flake on the waveguide facet in the spectral region of 720-735 nm, measured at 4K. The dotted white line indicates the edge of the few-layer region while the solid line marks the edges of the monolayer. Several areas of strong localized PL, corresponding to individual quantum dots, are present in the region surrounding the waveguide mode (marked by the dash-dot line). (b) Spatial map of localized PL from monolayer $WSe_2$ with individual single photon emitters marked by Roman numerals I-VII. Color bars show the normalized intensity.

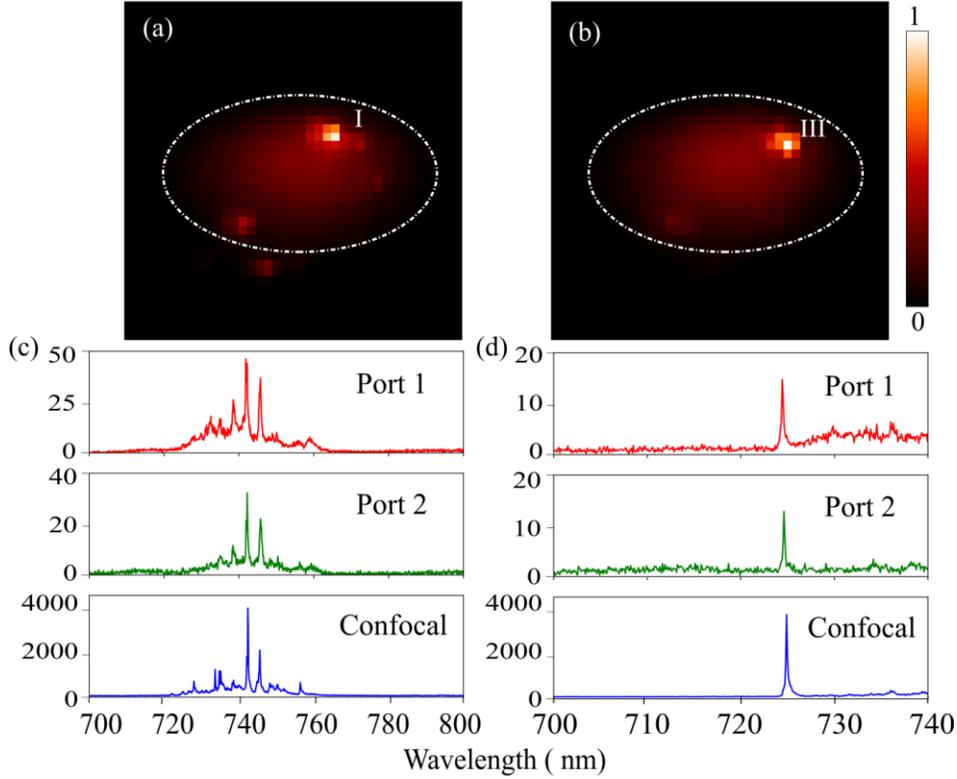

Fig. 3. (a) Spatially-resolved PL spectrally filtered at 745 nm and 725 nm (b). Color scale represents the normalized intensity. (c) PL spectra measured through the waveguide fiber output Ports 1 and 2 and confocally when the position of Emitter I is excited. (d) Spectra measured at Ports 1 and 2 and confocally when Emitter II is spatially excited. The spectra observed in confocal geometry is replicated at both output ports, albeit with a reduced photon count rate.

with NA = 0.07, in the ray optics approximation.

Using the transfer matrix approach, around 10% of emission from a dipole at an air-lithium niobate ($n = 2.2$) interface is collected into an objective of numerical aperture 0.82. Accounting for optical and fiber coupling losses, approximately 3% of the total emission is collected in the confocal geometry. In contrast, a fraction of 0.7% of the dipole emission is collected by a lithium niobate waveguide with numerical aperture NA = 0.07. Assuming 40% coupling efficiency from the waveguide into single-mode fibers, we expect 0.3% of the dipole emission at the fiber ports. Therefore, the expected ratio between the collection efficiency through the fibers (Ports 1 & 2) and through the confocal port is approximately 10% for a dipole located at the mode center.

We confirm these values through finite-difference time-domain (FDTD) simulations, modelling a dipole with varying displacement from the mode centre. For a dipole positioned at the mode centre we expect a collection efficiency of 0.3% (including fiber coupling losses), which reduces to 0.12% when the dipole is located 1.1 μm away from the mode centre. Taking the ratio of the counts experimentally measured through both fiber output ports and confocally, we achieve a collection efficiency of 0.11 ± 0.02% for an emitter at position I (Fig. 3 (a)), which agrees well with FDTD and transfer matrix simulations. The measured and simulated collection efficiencies are displayed in Table 1 for emitters I-VII (as shown in Fig. 3). The low overall coupling efficiency of the device is associated with the small index contrast of the Ti in-diffused region and, consequently, with the small numerical aperture (0.07) of the device. This problem can be addressed by using ridge waveguides etched in thin film lithium niobate-on-insulator (LNOI) [25]–[28], [32], [33] which feature much larger NA.

**Table 1. Comparison of measured and simulated collection efficiency as a function of emitter displacement**

| Emitter | Wavelength (nm) | Displacement (μm) | Collection efficiency (%) | Simulated efficiency (%) |
|---|---|---|---|---|
| I | 745.5 | 1.1 | 0.11±0.02 | 0.12 |
| II | 731.2 | 1.3 | 0.1±0.03 | 0.12 |
| III | 724.8 | 1.8 | 0.02±0.01 | 0.06 |
| IV | 733.4 | 2.5 | 0.02±0.01 | |
| V | 795 | 3.2 | 0.01±0.01 | |
| VI | 734.5 | 3.4 | 0.01±0.01 | |
| VII | 751.6 | 3.5 | 0.01±0.01 | |

## CONCLUSION

In conclusion, we have achieved the integration of quantum emitters in atomically thin $WSe_2$ onto a Ti in-diffused lithium niobate waveguide with a fiber coupled, on-chip beam-splitter. PL from the quantum dots was measured confocally at the waveguide facet and identical emission was collected through each output port of the device, demonstrating the operation of the on-chip Hanbury-Brown and Twiss interferometer. We expect the integration of two-dimensional materials into lithium niobate elements will be a key platform in the development of integrated quantum photonic circuits. In particular, the high refractive index contrast of thin film LNOI waveguides can provide improved optical coupling between the quantum dots and the waveguide and local strain-inducing features (such as nanopillars) can be incorporated on the facet edge. The hybrid quantum emitter-LNOI platform can be extended with in-situ strain tuning capability via the piezoelectric effect as well as access to passive and active optical elements, including grating couplers, compact integrated beam-splitters and ring resonators, and efficient reconfigurable interferometers.

## FUNDING

AP acknowledges funding from Australian Research Council Centre for Quantum Computation and Communication Technology CE170100012; Australian Research Council Discovery Early Career Researcher Award, Project No. DE140101700; RMIT University Vice-Chancellors Senior Research Fellowship. Work at Heriot-Watt is supported by the EPSRC (grant numbers EP/L015110/1 and EP/P029892/1) and the ERC (number 725920).

## ACKNOWLEDGEMENTS

This work was performed in part at the Micro Nano Research Facility (MNRF) at RMIT University, Melbourne, part of the ANFF, a company established under the National Collaborative Research Infrastructure Strategy to provide nano-and microfabrication facilities for Australia's researchers. BDG thanks the Royal Society for a Wolfson Merit Award and the Royal Academy of Engineering for a Chair in Emerging Technology.

## APPENDIX